\documentclass[aps,prb,superscriptaddress,reprint,showpacs]{revtex4-1}

\usepackage[english]{babel}
\usepackage[utf8x]{inputenc}
\usepackage[T1]{fontenc}
\usepackage{url}

\usepackage{graphicx}

\usepackage{amsmath} 
\usepackage{amssymb} 
\usepackage[detect-all]{siunitx}
\usepackage{bm}
\usepackage{mathtools}

\usepackage{color}

\usepackage{layouts}

\renewcommand{\d}{\mathrm{d}}

\begin{document}

\title{Accurate calculation of the transverse anisotropy in perpendicularly magnetized multilayers} 

\author{Felix B\"{u}ttner}
\thanks{These authors contributed equally to this work.}
\email{felixbuettner@gmail.com}
\affiliation{Institute of Physics, Johannes-Gutenberg-Universit\"at Mainz, Staudinger Weg 7, 55128 Mainz, Germany}
\affiliation{Graduate School Materials Science in Mainz, Staudinger Weg 9, 55128 Mainz, Germany}
\affiliation{Institut f\"ur Optik und Atomare Physik, Technische Universit\"at Berlin, Stra\ss e des 17. Juni 135, 10623 Berlin, Germany}
\author{Benjamin Kr\"uger}
\thanks{These authors contributed equally to this work.}
\email{bkrueger@uni-mainz.de}
\affiliation{Institute of Physics, Johannes-Gutenberg-Universit\"at Mainz, Staudinger Weg 7, 55128 Mainz, Germany}
\author{Stefan Eisebitt}
\affiliation{Institut f\"ur Optik und Atomare Physik, Technische Universit\"at Berlin, Stra\ss e des 17. Juni 135, 10623 Berlin, Germany}
\affiliation{Helmholtz-Zentrum Berlin f\"ur Materialien und Energie GmbH, Hahn-Meitner-Platz 1, 14109 Berlin, Germany}
\author{Mathias Kl\"aui}
\affiliation{Institute of Physics, Johannes-Gutenberg-Universit\"at Mainz, Staudinger Weg 7, 55128 Mainz, Germany}
\affiliation{Graduate School Materials Science in Mainz, Staudinger Weg 9, 55128 Mainz, Germany}

\date{February 27, 2015}

\begin{abstract}

The transverse anisotropy constant and the related D\"oring mass density are key parameters of the one-dimensional model to describe the motion of magnetic domain walls. So far, no general framework is available to determine these quantities from static characterizations such as magnetometry measurements. Here, we derive a universal analytical expression to calculate the transverse anisotropy constant for the important class of perpendicular magnetic multilayers. All the required input parameters of the model, such as the number of repeats, the thickness of a single magnetic layer, and the layer periodicity, as well as the effective perpendicular anisotropy, the saturation magnetization, and the static domain wall width are accessible by static sample characterizations. We apply our model to a widely used multilayer system and find that the effective transverse anisotropy constant is a factor 7 different from the when using the conventional approximations, showing the importance of using our analysis scheme.

\end{abstract}




\maketitle 

\section{Introduction}

Multilayers with perpendicular magnetic anisotropy (PMA) are widely used in research as well as in applications.\cite{buttner_magnetic_2013} A particular focus of the research today is the investigation of domain wall dynamics in such materials. Theoretically, the dynamics of domain walls is often described\cite{boulle_current-induced_2011} by the one-dimensional model.\cite{malozemoff_magnetic_1979} In this model, the tilt angle $\psi$ of the spins in the domain wall with respect to the domain wall plane is used as the conjugated momentum to the domain wall position $q$. The energy associated with a change of $\psi$ can be described by an effective uniaxial anisotropy, the so-called transverse anisotropy with anisotropy constant $K_\bot$. In general, this anisotropy acts as an energy reservoir that leads to domain wall quasiparticle behavior, such as domain wall inertia\cite{rhensius_imaging_2010} due to an effective domain wall mass.\cite{thomas_oscillatory_2006}

The transverse anisotropy constant is a key parameter to describe domain wall dynamics: it is proportional the to critical excitation strength (field or current) beyond which the domain wall becomes non-steady (Walker breakdown)\cite{boulle_current-induced_2011} and it is related to the inertia of the domain wall through the D\"oring mass density $m_D$ through the simple relation\cite{doring_uber_1948,malozemoff_magnetic_1979}
\begin{align}
m_D &= \frac{M_s^2(1+\alpha^2)}{K_\bot\gamma^2\Delta_0},\label{eq:app:md_final}
\end{align}
where $\alpha$ is the Gilbert damping, $\gamma=\SI{1.76e11}{A s / kg}$ is the gyromagnetic ratio, and $\Delta_0$ the static domain wall width.

The effective transverse anisotropy constant $K_\bot$ is not directly accessible from static sample characterization. Existing theories predict $K_\bot$ only for homogeneous magnetic materials and only in the limits of infinitely thick samples (bulk samples),\cite{malozemoff_magnetic_1979} where $K_\bot=1/2\mu_0M_s^2$, and for ultra-thin films of thickness $\mathcal{T}\rightarrow0$,\cite{tarasenko_bloch_1998} where $K_\bot = \ln(2)\mathcal{T}\mu_0M_s^2/(2\pi\Delta_0)$. As shown in this paper, these approximations yield significantly inaccurate results if applied to typical multilayer films, which typically are the application-relevant systems. 

In this paper, we present rigorous calculations of the magnetostatic energy of a multilayer system with a domain wall, from which we obtain analytical expressions for the transverse anisotropy constant $K_{\bot}$ as a function of the thickness of a single magnetic layer $\mathcal{T}$, the multilayer periodicity $\mathcal{P}$, the number of repeats $\mathcal{N}$, the static domain wall width $\Delta_0$, and the saturation magnetization $M_s$. We find that the saturation magnetization enters the equation only as a simple linear scaling factor and that $\Delta_0$, $\mathcal{P}$, and $\mathcal{T}$ enter only in the ratios $p := \mathcal{P}/(2\pi  \Delta_0)$, $t := \mathcal{T}/(2\pi  \Delta_0)$, and $\tau := \mathcal{P}/\mathcal{T}$. We provide accurate equations for $K_\bot$ as the main result of this paper in Eq.~\eqref{eq:app:sigma_d_G_Taylor_done} and we will show that
\begin{align}
\begin{alignedat}{1}K_\bot(t, p, \mathcal{N}) &\approx  \mu_0M_s^2
\frac{d^2f(\tau,\mathcal{N})+d\ln(2)}
{2d^2f(\tau,\mathcal{N})+d\,g(\tau,\mathcal{N})+1}
\end{alignedat}\label{eq:app:kperp_simplest_form}
\end{align}
with
\begin{align}
\begin{alignedat}{1}
f(\tau,\mathcal{N})&=\left(0.9\tau-0.76\right)/\mathcal{N}\\
g(\tau,\mathcal{N})&=\tau(1.8-0.05\ln(\mathcal{N}+12)+0.05\ln(\tau))\\
d&=t\mathcal{N}
\end{alignedat}
\end{align}
approximates the exact result with less than \SI{9}{\percent} error for $\num{e-4}\leq t\mathcal{N}\leq\num{e5}$, $1<\mathcal{N}\leq128$, and $1.25\leq \tau \leq 8$ and for $\tau=1$ and $\mathcal{N}=1$, which covers many systems that are intensely investigated today.


\section{General equations}

\begin{figure}
\centering
\includegraphics[width=\linewidth]{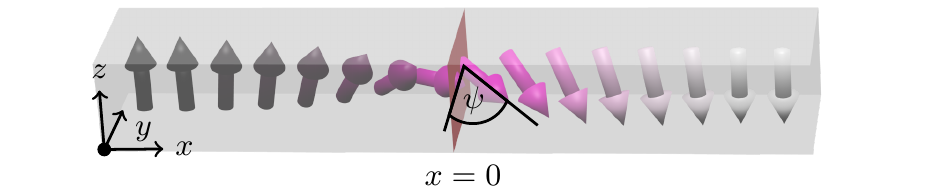}
\caption{Magnetization profile assumed by the calculations in this paper. The profile shows a domain wall at $x=0$, separating a domain where all magnetic moments point upwards (black arrows at the left side of the picture) from a domain where all magnetic moments point downwards (white arrows at the right side of the picture). The angle in the $x$-$y$-plane is the so-called transverse angle $\psi$. This angle is constant for all magnetic moments.}
\label{fig:magnetization_profile}
\end{figure}

The transverse anisotropy constant is a parameters of the one dimensional (1D) model. The following calculations are based on this model, which assumes the presence of a single, straight domain wall in the $y$-$z$-plane with the magnetization profile
\begin{align}
M_{x}(x) &= M_s \cosh^{-1}(x/\Delta) \sin(\psi)\label{eq:app:Bloch_Mx},\\
M_{y}(x) &= M_s \cosh^{-1}(x/\Delta) \cos(\psi)\label{eq:app:Bloch_My},\\
M_{z}(x) &= M_s \tanh(x/\Delta)\label{eq:app:Bloch_Mz},
\end{align}
as depicted in Fig.~\ref{fig:magnetization_profile}. Here, $\mathbf{M}=(M_{x},M_{y},M_{z})$ is the local magnetization, $M_s = |\mathbf{M}|$ is the saturation magnetization, $\Delta=\Delta(\psi)$ is the domain wall width parameter, and $\psi$ is the transverse (domain wall) angle. The magnetization is homogeneous along the wall, i.e., it does not depend on $y$ and $z$. The magnetostatic energy density $\sigma_{d}$ of such a domain wall (energy per unit area of the wall) is symmetric in $\psi$ and therefore can be described as a function of $s:=\sin^2\psi$. Also, the magnetostatic energy can be read as a uniaxial anisotropy with the hard axis along the $x$ direction. The general formula for such a uniaxial 
anisotropy reads
\begin{align}
\sigma_{d}(s) &= C + K_{\bot} \int\d x (M_x/M_s)^2 + \mathcal{O}(s^2)\\
 &= C + 2 \Delta_0 K_{\bot} s + \mathcal{O}(s^2),\label{eq:app:uniaxialAnisotropy}
\end{align}
where $C$ is a constant and $K_{\bot}$ is the leading order anisotropy constant. The first order partial derivative of Eq.~\eqref{eq:app:uniaxialAnisotropy} at $s=0$ yields $K_{\bot}$:
\begin{align}
K_\bot = \frac{\sigma_d^{(1)}}{2\Delta_0}\label{eq:app:Kperp},
\end{align} 
where the superscript $^{(1)}$ denotes the partial derivative with respect to $s$ at $s=0$. We will hence derive a formula to calculate the magnetostatic energy $\sigma_d^{(1)}$.

\section{Magnetostatic energy density}

In the following, we calculate the magnetostatic energy density $\sigma_d$ and, in particular, the first order partial derivative $\sigma_d^{(1)}$. We consider a thin film multilayer system of length $\mathcal{L}$ ($x$-direction) and width $\mathcal{W}$ ($y$-direction) in the limit $\mathcal{L},\mathcal{W}\rightarrow\infty$, which is a reasonably accurate approximation if $\mathcal{L},\mathcal{W}\gg\Delta_0$. Temporarily, we will treat for the calculation the width as a finite value and set it to infinity at a later stage. The ground state of such a perpendicular magnetic film is characterized by $s=0$. As in the 1D model, we will assume the presence of a single domain wall in the $y$-$z$-plane at $x=0$. However, now we have the additional $y$ dependency of the finite width and the $z$-dependency of the multilayer, which we describe as follows:
\begin{align}
M_{x}(x,z) &= M_{x}(x) w(y) \sum_{j=0}^{\mathcal{N}-1}v(z-j\mathcal{P}) \label{eq:app:Bloch_Mx_multilayer},\\
M_{y}(x,z) &= M_{y}(x) w(y) \sum_{j=0}^{\mathcal{N}-1}v(z-j\mathcal{P}) \label{eq:app:Bloch_My_multilayer},\\
M_{z}(x,z) &= M_{z}(x) w(y) \sum_{j=0}^{\mathcal{N}-1}v(z-j\mathcal{P}) \label{eq:app:Bloch_Mz_multilayer},\\
w(y) &= \theta\left(\mathcal{W}/2-|y-\mathcal{W}/2|\right),\\
v(z) &= \theta\left(\mathcal{T}/2-|z-\mathcal{T}/2|\right).
\end{align}
The magnetostatic energy of a magnetization configuration $\mathbf{M}(\mathbf{r})$ per unit domain wall area reads
\begin{align}
\sigma_d(s) &= \frac{\mu_0}{8\pi \mathcal{W}\mathcal{N}\mathcal{T}}\iint \d^3 \mathbf{r}\d^3 \mathbf{r}^\prime
\rho(\mathbf{r})\rho(\mathbf{r}^\prime)\frac{1}{|\mathbf{r}-\mathbf{r}^\prime|},
\label{eq:app:demagEnergyDensity}
\end{align}
where $\rho$ is the volume density of magnetic charges, $\mathcal{W}$ is the width of the structure ($y$-direction) and $\mathcal{N}\mathcal{T}$ is the total thickness of the magnetic material ($z$-direction). Magnetic charges arise from the divergence of the magnetization inside the volume (volume charges) and from the components of the magnetization perpendicular to the surfaces of the structure (surface charges). However, the surface charges do not explicitly depend on $s$ and the interactions between volume charges and surface charges vanish when averaged over $y$ and $z$ for symmetry reasons since all charges are antisymmetric around $\mathbf{r}=0$. Therefore, the only relevant charges are the volume charges
\begin{align}
\rho(\mathbf{r}) &= \sum_{j=0}^{\mathcal{N}-1} \rho_{1}(x,y,z-j\mathcal{P}) ,\\
\rho_{1}(\mathbf{r}) &= \rho_{1}(x) w(y)t(z), \\
\rho_{1}(x) &= -\operatorname{div}\mathbf{M} = \frac{M_s}{\Delta}\sin(\psi)\frac{\tanh(x/\Delta)}{\cosh(x/\Delta)}\label{eq:app:volume_charge_density}.
\end{align}
To solve the stray field integral Eq.~\eqref{eq:app:demagEnergyDensity}, we will first simplify the general form of the integral using the symmetry properties of the charge distribution Eq.~\eqref{eq:app:volume_charge_density} and use the explicit form of the charge distribution at a later stage. The kernel of the double sum in the integral
\begin{align}
 &
\begin{aligned}
  \sigma_{d} 
  &=  \frac{\mu_0}{8\pi \mathcal{W}\mathcal{N}\mathcal{T}}\iint \d^3 \mathbf{r}\d^3 \mathbf{r}^\prime \\
  &\times  \sum_{i,j=0}^{\mathcal{N}-1}
                \rho_1(x,y,z-i\mathcal{P}) \rho_1(x^\prime,,z^\prime-j\mathcal{P}) 
                \frac{1}{|\mathbf{r}-\mathbf{r}^\prime|} \\
\end{aligned}\\
&\begin{aligned}
  \hphantom{\sigma_{d}} 
  &= \frac{\mu_0}{8\pi \mathcal{W}\mathcal{N}\mathcal{T}}\iint \d^3 \mathbf{r}\d^3 \mathbf{r}^\prime
     \rho_1(x,y,z) \rho_1(x^\prime,y^\prime,z^\prime) \\
  &\times  \sum_{i,j=0}^{\mathcal{N}-1}
                \frac{1}{|\mathbf{r}+i\mathcal{P}\textbf{e}_{z}-\mathbf{r}^\prime-j\mathcal{P}\textbf{e}_{z}|}
\end{aligned}
\end{align}
depends on $i$ and $j$ only in the form $i-j$. That is, we can reorder the double sum to just one over this difference $(i-j)=(-\mathcal{N}+1),...,\mathcal{N}-1$ (which we will index again by $j$) and a factor $\mathcal{N}-|j|$ that counts how often the term $j$ is represented in the original sum:
\begin{align}
\begin{aligned}
\sigma_{d} &= 
\frac{\mu_0}{8\pi \mathcal{W}\mathcal{N}\mathcal{T}}\iint \d^3 \mathbf{r}\d^3 \mathbf{r}^\prime
\rho_1(\mathbf{r}) \rho_1(\mathbf{r}^\prime) \\
&\times\sum_{j=-\mathcal{N}+1}^{\mathcal{N}-1}
\frac{\mathcal{N}-|j|}{|\mathbf{r}-\mathbf{r}^\prime+j\mathcal{P}\textbf{e}_{z}|}.
\end{aligned}
\end{align}

The charge density $\rho$ has only a trivial dependence on $y$ and $z$. Hence, we can easily solve the integration with respect to $y$ and $y^\prime$, giving
\begin{align}
&\begin{alignedat}{1}\sigma_{d}  &=  \frac{\mu_0}{8\pi \mathcal{N}\mathcal{T}}\iint \d x \d x^\prime \int_0^{\mathcal{T}} \hspace{-.5em} \d z \int_0^{\mathcal{T}} \hspace{-.5em} \d z^\prime \rho_1(x) \rho_1(x^\prime) \\
&\times 
\sum_{j=-\mathcal{N}+1}^{\mathcal{N}-1} \hspace{-.5em}
(\mathcal{N}-|j|)\, f_{\mathcal{W}}(x-x^\prime,z-z^\prime+j\mathcal{P})
\end{alignedat}
\end{align}
with
\begin{align}
f_{\mathcal{W}}(x,z) &= \frac{1}{\mathcal{W}}\int_0^{\mathcal{W}} \hspace{-.5em} \d y \int_0^{\mathcal{W}} \hspace{-.5em} \d y^\prime 
\left[
\frac{1}{\sqrt{(y-y^\prime)^2 + x^2 + z^2}}
\right] \label{eq:app:f_w_before_integration}\\
&= -2+2\ln(2\mathcal{W})-\ln(x^2+z^2)+\mathcal{O}(\mathcal{W}^{-1}).
\end{align}

The steps to solve the integral Eq.~\eqref{eq:app:f_w_before_integration} are discussed in detail in appendix~\ref{app:sec:y-integration}. The terms constant in $x$ vanish with the integration over $x$ because the magnetic charges are antisymmetric in $x$. That is, in the limit $\mathcal{W}\rightarrow\infty$,
\begin{align}
\sigma_{d} &=  \frac{\mu_0}{8\pi \mathcal{N}\mathcal{T}}\sum_{j=-\mathcal{N}+1}^{\mathcal{N}-1}\hspace{-.5em}
(\mathcal{N}-|j|) \int_0^{\mathcal{T}} \hspace{-.5em}\d z \int_0^{\mathcal{T}} \hspace{-.5em}\d z^\prime  \iint \d x \d x^\prime  \nonumber\\
&\times 
\rho_1(x) \rho_1(x^\prime) h(x-x^\prime,z-z^\prime+j\mathcal{P})
\end{align}
with
\begin{align}
h(x,z) &= -\ln(x^2+z^2)
\end{align}
As outlined in appendix~\ref{app:sec:fourierSpaceIdentities}, the double integral over $x$ and $x^\prime$ can be reduced to a single integral in Fourier space. The result reads (the hat denotes a Fourier transform with respect to the $x$-coordinate):
\begin{align}
&\begin{alignedat}{1}\sigma_{d}^{(1)} &=  \frac{\mu_0M_s^2\sqrt{2\pi}}{8\pi \mathcal{N}\mathcal{T}}\sum_{j=-\mathcal{N}+1}^{\mathcal{N}-1}\hspace{-.5em}
(\mathcal{N}-|j|) \int_0^{\mathcal{T}}\hspace{-.5em} \d z \int_0^{\mathcal{T}} \hspace{-.5em} \d z^\prime   \label{eq:app:stray_field_energy_volume_h} \\
&\times 
 \int \d k |\hat{\varsigma}(k)|^2 \hat{h}(k,z-z^\prime+j\mathcal{P}),\end{alignedat}
\end{align}
with
\begin{align}
\varsigma(x) &= \frac{1}{\Delta_0}\frac{\tanh(x/\Delta_0)}{\cosh(x/\Delta_0)}, \\
\hat{\varsigma}(k) &= ik\Delta_0\sqrt{\frac{\pi}{2}}\frac{1}{\cosh\left(\frac{\pi\Delta_0k}{2}\right)}, \\
\hat{h}(k,z) &= \sqrt{2 \pi} \frac{1}{|k|} e^{-|z k|}.
\end{align}

The only term that depends on $z$ is the function $\hat{h}$. The integration of this part with respect to $z$ and $z^\prime$ is lengthy but not very involved, as shown in detail in appendix~\ref{app:sec:z-integration}. The result reads
\begin{align}
&\begin{alignedat}{1}\sigma_d^{(1)} &= \frac{\mu_0 M_s^2 \Delta_0}{8 t} \sum_{j=-\mathcal{N}+1}^{\mathcal{N}-1} \frac{\mathcal{N}-|j|}{\mathcal{N}} \sum_{i=-1}^1 (3|i|-2)\\
&\times \int_0^{\infty} \d q  \frac{e^{-|jp+it|q} + |jp+it|q - 1}{q \cosh^2(q/4)},
\end{alignedat}
\end{align}
where we have introduced the reduced variables $q:=2\pi\Delta_0k$, $p := \frac{\mathcal{P}}{2\pi \Delta_0}$, and $t := \frac{\mathcal{T}}{2\pi  \Delta_0}$. The $-1$ in the integral has been added to make each integral finite; it cancels out in the sum over $i$. We have furthermore replaced the integral over all $q$ by twice the integral over $q$ from $0$ to $\infty$. The integral

\begin{align}
G(x) &:=\int_0^\infty \d q \frac{e^{-xq}+xq-1}{q\cosh^2(q/4)}\label{eq:app:correct_but_complicated_solution}
\end{align}
can be solved analytically, as shown in appendix~\ref{app:sec:q-integration}. The final result reads
\begin{align}
&\begin{alignedat}{1}\sigma_d^{(1)} &= \frac{\mu_0 M_s^2 \Delta_0}{8 t} \sum_{j=-\mathcal{N}+1}^{\mathcal{N}-1} \frac{\mathcal{N}-|j|}{\mathcal{N}}\label{eq:app:sigma_d_G_Taylor_remains}\\
&\times ( G(|jp+t|) + G(|jp-t|) - 2 G(|jp|) ),
\end{alignedat}
\end{align}
where the function $G(x)$ is explicitly given in Eq.~\eqref{eq:app:g}. This result is exact but very lengthy and difficult to evaluate. We will optimize the result for numerical evaluation in the next section.

\section{Evaluation}

For practical purposes, the evaluation of Eq.~\eqref{eq:app:sigma_d_G_Taylor_remains} is often computationally too expensive. Also, we note that Eq.~\eqref{eq:app:sigma_d_G_Taylor_remains} contains differences of $G$, which means that a numerical evaluation of $G$ may yield dramatically wrong results just due to the finite precision of computer algebra and the error made cannot be estimated easily. We will therefore re-write Eq.~\eqref{eq:app:sigma_d_G_Taylor_remains} to eliminate the differences to make it more robust for numerical evaluation. In the following steps, we use that $jp \geq t$ for $j>0$ and that $G(0)=0$ to re-write the sum over j and to eliminate the absolute signs in the arguments of $G$. Subsequently, we expand $G(x)$ in a Taylor series around $jp$.
\begin{align}
&\sum_{j=-\mathcal{N}+1}^{\mathcal{N}-1} \frac{\mathcal{N}-|j|}{\mathcal{N}} ( G(|jp+t|) + G(|jp-t|) - 2 G(|jp|) ) \nonumber\\
&= 2 G(t) \\&\phantom{=}+ 2 \sum_{j=1}^{\mathcal{N}-1} \frac{\mathcal{N}-j}{\mathcal{N}} ( G(jp+t) + G(jp-t) - 2 G(jp) ) \nonumber\\
&= 2 G(t) + 4 \sum_{j=1}^{\mathcal{N}-1} \frac{\mathcal{N}-j}{\mathcal{N}} \sum_{m=1}^{\infty} \frac{t^{2m}}{(2m)!} G^{(2m)}(jp)\\
&= 2 G(t) + 4 \sum_{j=1}^{\mathcal{N}-1} \frac{\mathcal{N}-j}{\mathcal{N}} \sum_{m=1}^{\infty} \frac{t^{2m}}{(2m)!} F^{(2m-1)}(jp),\label{eq:app:sum_simplified}
\end{align}
where we have introduced the derivative of $G$:
\begin{align}
F(x) &:=\int_0^\infty \d q \frac{1-e^{-xq}}{\cosh^2(q/4)} \\
&= 8 x \left(H_x-H_{x-\frac{1}{2}}\right),
\end{align}
where $H_x$ is the harmonic number (see appendix~\ref{app:sec:q-integration} for the steps to solve the integral). The Taylor series converges for all $t,p$, and $\mathcal{N}$ and the derivatives $F^{(n)}(x)$ are easy to calculate and to evaluate. It is useful to note that $F^{(n)}(x)$ are all positive (negative) for odd (even) $n$, finite for $x\in[0,\infty]$, and monotonically approaching $0$ as $x\rightarrow\infty$. We estimate the error by the Lagrange error bound for Taylor series, which states that the error $E_n$ by approximating a function $f$ at a position $x$ using a Taylor series of order $n$ around $x_0$ has an upper bound of
\begin{align}
E_n \leq \frac{\underset{x\in[x_0,x]}{\text{max}}|f^{(n+1)}(x)|}{(n+1)!}|x-x_0|^{(n+1)}.\label{eq:app:Lagrange_error}
\end{align}
Inserting this back, we obtain the formula for approximating $\sigma_d^{(1)}$ up to order $M$
\begin{align}
\sigma_d^{(1)}(&M) = \frac{\mu_0 M_s^2 \Delta_0}{4 t} \label{eq:app:sigma_d_G_Taylor_done}\\\nonumber
&\times \left(G(t) + 2 \sum_{j=1}^{\mathcal{N}-1} \frac{\mathcal{N}-j}{\mathcal{N}} \sum_{m=1}^{M} \frac{t^{2m}}{(2m)!} F^{(2m-1)}(jp)\right)
\end{align}
with the error
\begin{align}
\Delta\sigma_d^{(1)}(M) &= \frac{\mu_0 M_s^2 \Delta_0}{2 t}\label{eq:app:error_sigma_d_G_Taylor_done}\\\nonumber
&\times \sum_{j=1}^{\mathcal{N}-1} \frac{\mathcal{N}-j}{\mathcal{N}} \frac{t^{2M+1}}{(2M+1)!} |F^{(2M)}(jp-t)| 
\end{align}
where we have used that $F^{(2m)}$ is monotonic to derive the maximum that enters Eq.~\eqref{eq:app:Lagrange_error}.

\begin{figure}
\centering
\includegraphics[width=\linewidth]{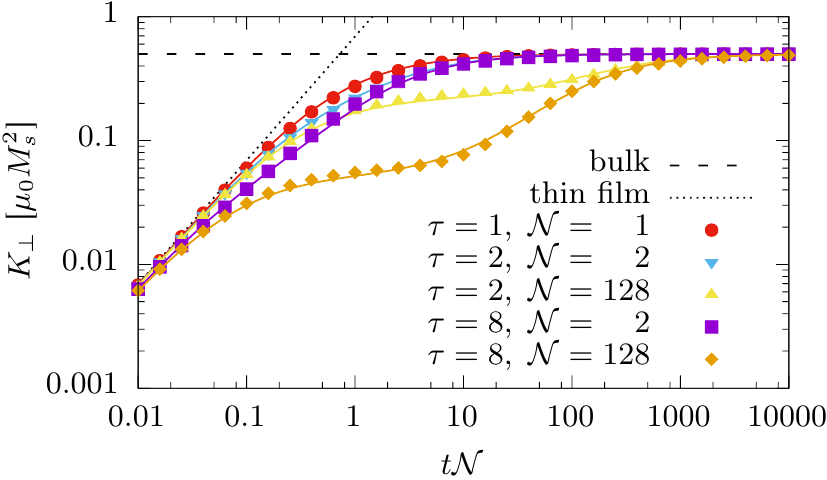}
\caption{Transverse anisotropy constant $K_\bot$ as a function of the reduced single layer thickness $t$, the reduced layer periodicity $p$, and the total number of layers $\mathcal{N}$ (points). The solid lines show the approximations according to Eq.~\eqref{eq:app:kperp_simplest_form}. The dashed line shows the approximation of a very thick (bulk) sample and the dotted line indicates the results of the thin film approximation. Both approximations are significantly inaccurate in the technologically relevant region of multilayers of intermediate thickness.}
\label{fig:kperp}
\end{figure}

\begin{figure}
\centering
\includegraphics[width=\linewidth]{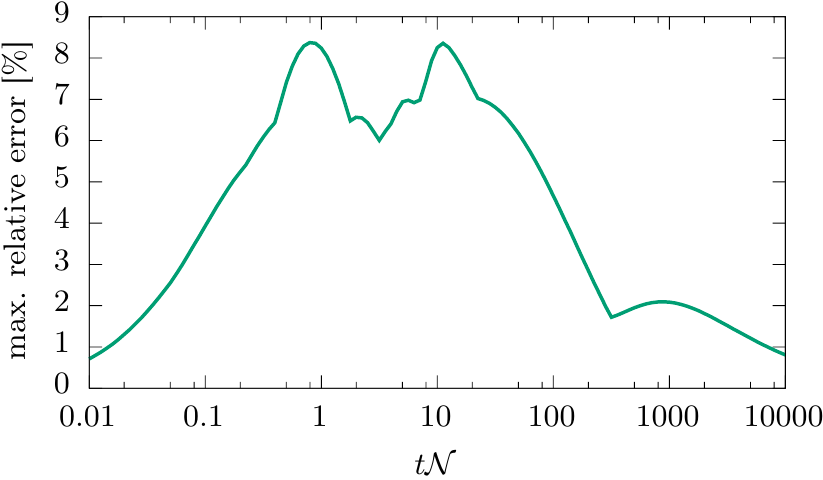}
\caption{Maximum error made by the approximation of Eq.~\eqref{eq:app:kperp_simplest_form}, compared to the accurate result of Eq.~\eqref{eq:app:sigma_d_G_Taylor_done}, as a function of $t\mathcal{N}$ for $\mathcal{N}$ in the range of \numrange{1}{128} and $\tau$ in the range of \numrange{1}{8}.}
\label{fig:kperp_error}
\end{figure}

From Eqs.~\eqref{eq:app:Kperp} and \eqref{eq:app:sigma_d_G_Taylor_done} we can now derive the transverse anisotropy constant. For the general case of variable $\mathcal{N}$ and arbitrary $t$ and $p$, the anisotropy constant $K_\bot$ is plotted in Fig.~\ref{fig:kperp}. We find that this hyper-dimensional dependency can be very well approximated by the simple rational form given in Eq.~\eqref{eq:app:kperp_simplest_form}.
The error made by this approximation is plotted in Fig.~\ref{fig:kperp_error} for $\log_{10}(t\mathcal{N})$ in the range of \numrange{-4}{5} in steps of \num{0.05}, $\log_{2}\mathcal{N}$ in the range of \numrange{0}{7} in unit steps and $\tau$ in the range of \numrange{1}{8} in steps of \num{0.25}. The error is always smaller than $\SI{8}{\percent}$. That is, for most multilayer systems investigated today, Eq.~\eqref{eq:app:kperp_simplest_form} provides a very good description of the transverse anisotropy constant.

\section{Applications}

We first note that, for a single homogeneous magnetic film, Eq.~\eqref{eq:app:sigma_d_G_Taylor_done} simplifies to
\begin{align}
K_\bot(\mathcal{N}=1) = \frac{\mu_0M_s^2}{4t}G(t),
\end{align}
which converges to $K_\bot = \mu_0M_s^2/2$ for a bulk material ($t=\infty$) and to $K_\bot = \ln(2)t\mu_0M_s^2$ for a very thin film ($t\rightarrow0$), as expected.

The most important application of our calculations is the case of a true multilayer system, i.e., for $N>1$. An example is a Pt(2)/[Co$_{68}$B$_{32}$(0.4)/Pt(0.7)]$_{30}$/ Pt(1.3) (thickness in nm) magnetic multilayer with $M_s=\SI{1.19 +- 0.03 e6}{A/m}$ and $\Delta_0=\SI{11 +- 2}{nm}$, which has been used in the investigation of domain dynamics.\cite{buttner_magnetic_2013,buttner_dynamics_2015} Entering this into our calculation, we find that the transverse anisotropy constant is given by $0.07(2) \mu_0 M_s^2$. This value is a factor seven smaller than predicted by the simple and widely used formula $K_\bot=\mu_0M_s^2/2$ and a factor 1.7 smaller than predicted by the thin film formula $K_\bot = \ln(2)t\mu_0M_s^2$. This means that quantitative predictions of the domain wall dynamics, such as the Walker breakdown,\cite{boulle_current-induced_2011} are inaccurate when relying on existing formulas, showing that one needs to take into account our calculations for a realistic calculation of the resulting domain wall dynamics.

\section{Conclusions}

In conclusion, we have derived an accurate analytical expression for the transverse anisotropy constant in perpendicular magnetic multilayer films as used in a 1D model description. The transverse anisotropy constant can be reliably computed using magnetic properties that can be ascertained easily for multilayer films from static measurements. We have shown that the results can significantly deviate from existing and commonly used theories that oversimplify the actual sample configuration and are only valid in the very thin or very thick limit but not in the intermediate regime, which is however most widely used for devices. We have provided a more accurate simplification yielding sufficiently precise results for most systems investigated today that are easily accessible without very involved computation. In particular we show that for a commonly used multilayer stack our result describes the anisotropy well compared to the conventionally used approximations, which are off by up to a factor 7. Our results enable precise modeling of domain wall dynamics in systems with perpendicular magnetic anisotropy using the 1D model, which has previously failed due to inaccurate assumptions of the transverse anisotropy.

\begin{acknowledgments}

This work was funded by the German Ministry for Education and Science (BMBF) through the projects MULTIMAG (13N9911) and MPSCATT (05K10KTB), EU’s 7th Framework Programme MAGWIRE (FP7-ICT-2009-5 257707) and WALL (FP7-PEOPLE-2013-ITN), the European Research Council through the Starting Independent Researcher Grant MASPIC (ERC-2007-StG 208162), the Mainz Center for Complex Materials (COMATT), the Graduate School Materials Science in Mainz and the Deutsche Forschungsgemeinschaft (DFG). BK is grateful for financial support by the Carl Zeiss Stiftung.

\end{acknowledgments}

\appendix

\section{$y$-integration}
\label{app:sec:y-integration}

Here, we will calculate the integral 
\begin{align}
\frac{1}{\mathcal{W}}\int_0^{\mathcal{W}} \d y \int_0^{\mathcal{W}} \d y^\prime \frac{1}{\sqrt{(y-y^\prime)^2+a}}
\end{align}
for large $\mathcal{W}$. The integration with respect to $y^\prime$ yields
\begin{align}
&\begin{alignedat}{1}&\int_0^{\mathcal{W}} \d y^\prime \frac{1}{\sqrt{(y-y^\prime)^2+a}}\end{alignedat} \\ 
&\begin{alignedat}{2}
&= &&-\left.\ln\left(\sqrt{(y-y^\prime)^2+a}+y-y^\prime\right)\right|_{y^\prime=0}^{\mathcal{W}} \end{alignedat}\\
&\begin{alignedat}{1}&= -f(y-\mathcal{W})+f(y) \end{alignedat},\\
&\begin{alignedat}{1}f(y) &= \ln\left(\sqrt{y^2+a}+y\right).\end{alignedat}
\end{align}
The antiderivative of $f$ reads
\begin{align}
F(y)=y\ln\left(\sqrt{a+y^2}+y\right)-\sqrt{a+y^2}
\end{align}
and the total integral is therefore given by
\begin{align}
&\begin{alignedat}{1}&\frac{1}{\mathcal{W}}\int_0^{\mathcal{W}} \d y \int_0^{\mathcal{W}} \d y^\prime \frac{1}{\sqrt{(y-y^\prime)^2+a^2}} \\
&= (F(\mathcal{W})+F(-\mathcal{W})-2F(0))/\mathcal{W} \end{alignedat}\\
&\begin{alignedat}{2}&= &&\ln\left(\sqrt{a/\mathcal{W}^2+1}+1\right)-\ln\left(\sqrt{a/\mathcal{W}^2+1}-1\right) \\
& &&-2\sqrt{a/\mathcal{W}^2+1}+2\sqrt{a/\mathcal{W}^2} \end{alignedat}\\
&\begin{alignedat}{2}&= &&-2+2\ln(2\mathcal{W})-\ln(a)+\mathcal{O}(\mathcal{W}^{-1}). \end{alignedat}
\end{align}

\section{Fourier space identities}
\label{app:sec:fourierSpaceIdentities}

Here, we show how to reduce a typical double integral over a pair interaction kernel to a single integral in Fourier space (assuming real-valued functions):
\begin{align}
&\iint \d x \d x^\prime f(x) h(x^\prime) g(x-x^\prime) \\\nonumber
&= \int \d x  f(x) \int \d x^\prime g(x-x^\prime) h(x^\prime) \\
 &= \int \d x  f(x) (h*g)(x) \\
 &= \int \d x  f(x)  \int \d k \hat h(k) \hat g(k) e^{-ikx} \\
 &= \sqrt{2\pi}\int \d k \hat h(k) \hat g(k) \frac{1}{\sqrt{2\pi}} \int \d x  f(x) e^{-ikx} \\
 &= \sqrt{2\pi}\int \d k \hat h(k) \hat g(k) \left(\hat f(k)\right)^*
\end{align}
Note that we are using the following definition of the Fourier transform:
\begin{align}
\hat f(k) &= \frac{1}{\sqrt{2\pi}}\int \d x f(x) e^{ikx}\\
f(x) &= \frac{1}{\sqrt{2\pi}}\int \d k \hat f(k) e^{-ikx}
\end{align}

\section{$z$-integration}
\label{app:sec:z-integration}

In the following we show how to perform the integration of Eq.~\eqref{eq:app:stray_field_energy_volume_h}. We first note that
\begin{align}
\iint e^{-k|z|}\d z = k^{-2}\left(e^{-k|z|}-k|z|\right)
\end{align}
is a continuous second antiderivative of $\exp(-k|z|)$. Furthermore, remember that
\begin{align}
&\int_a^b \d z \int_c^d \d z^\prime f(z^\prime - z) \\\nonumber
&= G(c-b)+G(d-a)-G(c-a)-G(c-b)
\end{align}
if $\frac{\d^2 G(z)}{\d z^2} = f(z)$. Using these identities, we see that

\begin{align}
&\int_0^{\mathcal{T}} \d z \int_0^{\mathcal{T}}  \d z^\prime\, \hat{h}(k,z-z^\prime+j\mathcal{P}) \\
&= \frac{\sqrt{2\pi}}{|k|}\int_0^{\mathcal{T}} \d z \int_0^{\mathcal{T}}  \d z^\prime e^{-|z-z^\prime+j\mathcal{P}| |k|} \\
&= \frac{\sqrt{2\pi}}{|k|}\int_0^{\mathcal{T}} \d z \int_{-j\mathcal{P}}^{\mathcal{T}-j\mathcal{P}}  \d z^\prime e^{-|z^\prime-z| |k|} \\
&= \frac{\sqrt{2 \pi}}{|k|^3} \sum_{i=-1}^1 (3|i|-2) \left( e^{-|j\mathcal{P}+i\mathcal{T}||k|} + |j\mathcal{P}+i\mathcal{T}||k| \right).
\end{align}

\section{$q$-integration}
\label{app:sec:q-integration}

In this section we provide the exact solution of the Fourier space integral Eq.~\eqref{eq:app:correct_but_complicated_solution}:
\begin{align}
\int_0^\infty \d q \frac{e^{-x q}+xq-1}{q \cosh^2(q/4)} &=: G(x).
\end{align}
To solve this, we note that
\begin{align}
G(x)=\int_0^x F(x^\prime)\d x^\prime
\end{align}
where the integral
\begin{align}
F(x) &=  \int_0^\infty \d q \frac{1-e^{-x q}}{\cosh^2(q/4)} \\
&= 4-4\int_0^\infty \d q \underbrace{e^{-x q}}_{f(q)}\underbrace{\frac{e^{q/2}}{(1+e^{q/2})^2}}_{g^\prime(q)}\\
&=4\int_0^\infty \d q (-x)e^{-x q}\frac{-2}{1+e^{q/2}}\\
&=8x\int_0^\infty \d q e^{-x q}\frac{e^{q/2}-1}{(e^{q/2}-1)(e^{q/2}+1)}\\
&= 8 x  \int_0^\infty \d q \frac{e^{-(x-1/2) q}-e^{-x q}}{e^{q}-1} \\
&= 8 x \int \d y \frac{y^{x-1/2}-y^x}{1-y} \\
&= 8 x \left(H_x-H_{x-\frac{1}{2}}\right)
\end{align}
can be solved through integration by parts. Here, $H$ is the harmonic number. We therefore obtain
\begin{align}
\begin{alignedat}{1}
G(x) &= -8 \bigg(\psi ^{(-2)}(x+1)-\psi ^{(-2)}\left(x+\frac{1}{2}\right) \\
&-s \ln (\Gamma (x+1))+x \ln \left(\Gamma \left(x+\frac{1}{2}\right)\right) \\
&-\psi ^{(-2)}(1)+\psi ^{(-2)}\left(\frac{1}{2}\right)\bigg),
\end{alignedat}\label{eq:app:g}
\end{align}
where $\gamma$ is gamma function and $\psi$ is the  digamma function.


\begin{thebibliography}{8}%
\makeatletter
\providecommand \@ifxundefined [1]{%
 \@ifx{#1\undefined}
}%
\providecommand \@ifnum [1]{%
 \ifnum #1\expandafter \@firstoftwo
 \else \expandafter \@secondoftwo
 \fi
}%
\providecommand \@ifx [1]{%
 \ifx #1\expandafter \@firstoftwo
 \else \expandafter \@secondoftwo
 \fi
}%
\providecommand \natexlab [1]{#1}%
\providecommand \enquote  [1]{``#1''}%
\providecommand \bibnamefont  [1]{#1}%
\providecommand \bibfnamefont [1]{#1}%
\providecommand \citenamefont [1]{#1}%
\providecommand \href@noop [0]{\@secondoftwo}%
\providecommand \href [0]{\begingroup \@sanitize@url \@href}%
\providecommand \@href[1]{\@@startlink{#1}\@@href}%
\providecommand \@@href[1]{\endgroup#1\@@endlink}%
\providecommand \@sanitize@url [0]{\catcode `\\12\catcode `\$12\catcode
  `\&12\catcode `\#12\catcode `\^12\catcode `\_12\catcode `\%12\relax}%
\providecommand \@@startlink[1]{}%
\providecommand \@@endlink[0]{}%
\providecommand \url  [0]{\begingroup\@sanitize@url \@url }%
\providecommand \@url [1]{\endgroup\@href {#1}{\urlprefix }}%
\providecommand \urlprefix  [0]{URL }%
\providecommand \Eprint [0]{\href }%
\providecommand \doibase [0]{http://dx.doi.org/}%
\providecommand \selectlanguage [0]{\@gobble}%
\providecommand \bibinfo  [0]{\@secondoftwo}%
\providecommand \bibfield  [0]{\@secondoftwo}%
\providecommand \translation [1]{[#1]}%
\providecommand \BibitemOpen [0]{}%
\providecommand \bibitemStop [0]{}%
\providecommand \bibitemNoStop [0]{.\EOS\space}%
\providecommand \EOS [0]{\spacefactor3000\relax}%
\providecommand \BibitemShut  [1]{\csname bibitem#1\endcsname}%
\let\auto@bib@innerbib\@empty
\bibitem [{\citenamefont {B\"{u}ttner}\ \emph {et~al.}(2013)\citenamefont
  {B\"{u}ttner}, \citenamefont {Moutafis}, \citenamefont {Bisig}, \citenamefont
  {Wohlh\"{u}ter}, \citenamefont {G\"{u}nther}, \citenamefont {Mohanty},
  \citenamefont {Geilhufe}, \citenamefont {Schneider}, \citenamefont
  {Korff~Schmising}, \citenamefont {Schaffert}, \citenamefont {Pfau},
  \citenamefont {Hantschmann}, \citenamefont {Riemeier}, \citenamefont {Emmel},
  \citenamefont {Finizio}, \citenamefont {Jakob}, \citenamefont {Weigand},
  \citenamefont {Rhensius}, \citenamefont {Franken}, \citenamefont {Lavrijsen},
  \citenamefont {Swagten}, \citenamefont {Stoll}, \citenamefont {Eisebitt},\
  and\ \citenamefont {Kl\"{a}ui}}]{buttner_magnetic_2013}%
  \BibitemOpen
  \bibfield  {author} {\bibinfo {author} {\bibfnamefont {F.}~\bibnamefont
  {B\"{u}ttner}}, \bibinfo {author} {\bibfnamefont {C.}~\bibnamefont
  {Moutafis}}, \bibinfo {author} {\bibfnamefont {A.}~\bibnamefont {Bisig}},
  \bibinfo {author} {\bibfnamefont {P.}~\bibnamefont {Wohlh\"{u}ter}}, \bibinfo
  {author} {\bibfnamefont {C.~M.}\ \bibnamefont {G\"{u}nther}}, \bibinfo
  {author} {\bibfnamefont {J.}~\bibnamefont {Mohanty}}, \bibinfo {author}
  {\bibfnamefont {J.}~\bibnamefont {Geilhufe}}, \bibinfo {author}
  {\bibfnamefont {M.}~\bibnamefont {Schneider}}, \bibinfo {author}
  {\bibfnamefont {C.~v.}\ \bibnamefont {Korff~Schmising}}, \bibinfo {author}
  {\bibfnamefont {S.}~\bibnamefont {Schaffert}}, \bibinfo {author}
  {\bibfnamefont {B.}~\bibnamefont {Pfau}}, \bibinfo {author} {\bibfnamefont
  {M.}~\bibnamefont {Hantschmann}}, \bibinfo {author} {\bibfnamefont
  {M.}~\bibnamefont {Riemeier}}, \bibinfo {author} {\bibfnamefont
  {M.}~\bibnamefont {Emmel}}, \bibinfo {author} {\bibfnamefont
  {S.}~\bibnamefont {Finizio}}, \bibinfo {author} {\bibfnamefont
  {G.}~\bibnamefont {Jakob}}, \bibinfo {author} {\bibfnamefont
  {M.}~\bibnamefont {Weigand}}, \bibinfo {author} {\bibfnamefont
  {J.}~\bibnamefont {Rhensius}}, \bibinfo {author} {\bibfnamefont {J.~H.}\
  \bibnamefont {Franken}}, \bibinfo {author} {\bibfnamefont {R.}~\bibnamefont
  {Lavrijsen}}, \bibinfo {author} {\bibfnamefont {H.~J.~M.}\ \bibnamefont
  {Swagten}}, \bibinfo {author} {\bibfnamefont {H.}~\bibnamefont {Stoll}},
  \bibinfo {author} {\bibfnamefont {S.}~\bibnamefont {Eisebitt}}, \ and\
  \bibinfo {author} {\bibfnamefont {M.}~\bibnamefont {Kl\"{a}ui}},\ }\href
  {\doibase 10.1103/PhysRevB.87.134422} {\bibfield  {journal} {\bibinfo
  {journal} {Physical Review B}\ }\textbf {\bibinfo {volume} {87}},\ \bibinfo
  {pages} {134422} (\bibinfo {year} {2013})}\BibitemShut {NoStop}%
\bibitem [{\citenamefont {Boulle}\ \emph {et~al.}(2011)\citenamefont {Boulle},
  \citenamefont {Malinowski},\ and\ \citenamefont
  {Kl\"{a}ui}}]{boulle_current-induced_2011}%
  \BibitemOpen
  \bibfield  {author} {\bibinfo {author} {\bibfnamefont {O.}~\bibnamefont
  {Boulle}}, \bibinfo {author} {\bibfnamefont {G.}~\bibnamefont {Malinowski}},
  \ and\ \bibinfo {author} {\bibfnamefont {M.}~\bibnamefont {Kl\"{a}ui}},\
  }\href {\doibase 10.1016/j.mser.2011.04.001} {\bibfield  {journal} {\bibinfo
  {journal} {Materials Science and Engineering: R: Reports}\ }\textbf {\bibinfo
  {volume} {72}},\ \bibinfo {pages} {159} (\bibinfo {year} {2011})}\BibitemShut
  {NoStop}%
\bibitem [{\citenamefont {Malozemoff}\ and\ \citenamefont
  {Slonczewski}(1979)}]{malozemoff_magnetic_1979}%
  \BibitemOpen
  \bibfield  {author} {\bibinfo {author} {\bibfnamefont {A.~P.}\ \bibnamefont
  {Malozemoff}}\ and\ \bibinfo {author} {\bibfnamefont {J.~C.}\ \bibnamefont
  {Slonczewski}},\ }\href@noop {} {\emph {\bibinfo {title} {Magnetic Domain
  Walls in Bubble Materials}}}\ (\bibinfo  {publisher} {Academic Press},\
  \bibinfo {address} {New York},\ \bibinfo {year} {1979})\BibitemShut {NoStop}%
\bibitem [{\citenamefont {Rhensius}\ \emph {et~al.}(2010)\citenamefont
  {Rhensius}, \citenamefont {Heyne}, \citenamefont {Backes}, \citenamefont
  {Krzyk}, \citenamefont {Heyderman}, \citenamefont {Joly}, \citenamefont
  {Nolting},\ and\ \citenamefont {Kl\"{a}ui}}]{rhensius_imaging_2010}%
  \BibitemOpen
  \bibfield  {author} {\bibinfo {author} {\bibfnamefont {J.}~\bibnamefont
  {Rhensius}}, \bibinfo {author} {\bibfnamefont {L.}~\bibnamefont {Heyne}},
  \bibinfo {author} {\bibfnamefont {D.}~\bibnamefont {Backes}}, \bibinfo
  {author} {\bibfnamefont {S.}~\bibnamefont {Krzyk}}, \bibinfo {author}
  {\bibfnamefont {L.~J.}\ \bibnamefont {Heyderman}}, \bibinfo {author}
  {\bibfnamefont {L.}~\bibnamefont {Joly}}, \bibinfo {author} {\bibfnamefont
  {F.}~\bibnamefont {Nolting}}, \ and\ \bibinfo {author} {\bibfnamefont
  {M.}~\bibnamefont {Kl\"{a}ui}},\ }\href {\doibase
  10.1103/PhysRevLett.104.067201} {\bibfield  {journal} {\bibinfo  {journal}
  {Physical Review Letters}\ }\textbf {\bibinfo {volume} {104}},\ \bibinfo
  {pages} {067201} (\bibinfo {year} {2010})}\BibitemShut {NoStop}%
\bibitem [{\citenamefont {Thomas}\ \emph {et~al.}(2006)\citenamefont {Thomas},
  \citenamefont {Hayashi}, \citenamefont {Jiang}, \citenamefont {Moriya},
  \citenamefont {Rettner},\ and\ \citenamefont
  {Parkin}}]{thomas_oscillatory_2006}%
  \BibitemOpen
  \bibfield  {author} {\bibinfo {author} {\bibfnamefont {L.}~\bibnamefont
  {Thomas}}, \bibinfo {author} {\bibfnamefont {M.}~\bibnamefont {Hayashi}},
  \bibinfo {author} {\bibfnamefont {X.}~\bibnamefont {Jiang}}, \bibinfo
  {author} {\bibfnamefont {R.}~\bibnamefont {Moriya}}, \bibinfo {author}
  {\bibfnamefont {C.}~\bibnamefont {Rettner}}, \ and\ \bibinfo {author}
  {\bibfnamefont {S.~S.~P.}\ \bibnamefont {Parkin}},\ }\href {\doibase
  10.1038/nature05093} {\bibfield  {journal} {\bibinfo  {journal} {Nature}\
  }\textbf {\bibinfo {volume} {443}},\ \bibinfo {pages} {197} (\bibinfo {year}
  {2006})}\BibitemShut {NoStop}%
\bibitem [{\citenamefont {D\"{o}ring}(1948)}]{doring_uber_1948}%
  \BibitemOpen
  \bibfield  {author} {\bibinfo {author} {\bibfnamefont {W.}~\bibnamefont
  {D\"{o}ring}},\ }\href@noop {} {\bibfield  {journal} {\bibinfo  {journal} {Z.
  Naturforsch. A}\ }\textbf {\bibinfo {volume} {3}},\ \bibinfo {pages} {373}
  (\bibinfo {year} {1948})}\BibitemShut {NoStop}%
\bibitem [{\citenamefont {Tarasenko}\ \emph {et~al.}(1998)\citenamefont
  {Tarasenko}, \citenamefont {Stankiewicz}, \citenamefont {Tarasenko},\ and\
  \citenamefont {Ferr\'{e}}}]{tarasenko_bloch_1998}%
  \BibitemOpen
  \bibfield  {author} {\bibinfo {author} {\bibfnamefont {S.~V.}\ \bibnamefont
  {Tarasenko}}, \bibinfo {author} {\bibfnamefont {A.}~\bibnamefont
  {Stankiewicz}}, \bibinfo {author} {\bibfnamefont {V.~V.}\ \bibnamefont
  {Tarasenko}}, \ and\ \bibinfo {author} {\bibfnamefont {J.}~\bibnamefont
  {Ferr\'{e}}},\ }\href {\doibase 10.1016/S0304-8853(98)00230-3} {\bibfield
  {journal} {\bibinfo  {journal} {Journal of Magnetism and Magnetic Materials}\
  }\textbf {\bibinfo {volume} {189}},\ \bibinfo {pages} {19} (\bibinfo {year}
  {1998})}\BibitemShut {NoStop}%
\bibitem [{\citenamefont {B\"{u}ttner}\ \emph {et~al.}(2015)\citenamefont
  {B\"{u}ttner}, \citenamefont {Moutafis}, \citenamefont {Schneider},
  \citenamefont {Kr\"{u}ger}, \citenamefont {G\"{u}nther}, \citenamefont
  {Geilhufe}, \citenamefont {Schmising}, \citenamefont {Mohanty}, \citenamefont
  {Pfau}, \citenamefont {Schaffert}, \citenamefont {Bisig}, \citenamefont
  {Foerster}, \citenamefont {Schulz}, \citenamefont {Vaz}, \citenamefont
  {Franken}, \citenamefont {Swagten}, \citenamefont {Kl\"{a}ui},\ and\
  \citenamefont {Eisebitt}}]{buttner_dynamics_2015}%
  \BibitemOpen
  \bibfield  {author} {\bibinfo {author} {\bibfnamefont {F.}~\bibnamefont
  {B\"{u}ttner}}, \bibinfo {author} {\bibfnamefont {C.}~\bibnamefont
  {Moutafis}}, \bibinfo {author} {\bibfnamefont {M.}~\bibnamefont {Schneider}},
  \bibinfo {author} {\bibfnamefont {B.}~\bibnamefont {Kr\"{u}ger}}, \bibinfo
  {author} {\bibfnamefont {C.~M.}\ \bibnamefont {G\"{u}nther}}, \bibinfo
  {author} {\bibfnamefont {J.}~\bibnamefont {Geilhufe}}, \bibinfo {author}
  {\bibfnamefont {C.~v.~K.}\ \bibnamefont {Schmising}}, \bibinfo {author}
  {\bibfnamefont {J.}~\bibnamefont {Mohanty}}, \bibinfo {author} {\bibfnamefont
  {B.}~\bibnamefont {Pfau}}, \bibinfo {author} {\bibfnamefont {S.}~\bibnamefont
  {Schaffert}}, \bibinfo {author} {\bibfnamefont {A.}~\bibnamefont {Bisig}},
  \bibinfo {author} {\bibfnamefont {M.}~\bibnamefont {Foerster}}, \bibinfo
  {author} {\bibfnamefont {T.}~\bibnamefont {Schulz}}, \bibinfo {author}
  {\bibfnamefont {C.~a.~F.}\ \bibnamefont {Vaz}}, \bibinfo {author}
  {\bibfnamefont {J.~H.}\ \bibnamefont {Franken}}, \bibinfo {author}
  {\bibfnamefont {H.~J.~M.}\ \bibnamefont {Swagten}}, \bibinfo {author}
  {\bibfnamefont {M.}~\bibnamefont {Kl\"{a}ui}}, \ and\ \bibinfo {author}
  {\bibfnamefont {S.}~\bibnamefont {Eisebitt}},\ }\href {\doibase
  10.1038/nphys3234} {\bibfield  {journal} {\bibinfo  {journal} {Nature
  Physics}\ }\textbf {\bibinfo {volume} {advance online publication}} (\bibinfo
  {year} {2015}),\ 10.1038/nphys3234}\BibitemShut {NoStop}%
\end{thebibliography}
\end{document}